# Interaction enhanced inter-site hoppings for holons and interlayer exciton insulators in moiré correlated insulators

Zijian Ma[1], Hongyi Yu[1,2*]

[1] Guangdong Provincial Key Laboratory of Quantum Metrology and Sensing & School of Physics and Astronomy, Sun Yat-Sen University (Zhuhai Campus), Zhuhai 519082, China

[2] State Key Laboratory of Optoelectronic Materials and Technologies, Sun Yat-Sen University (Guangzhou Campus), Guangzhou 510275, China

* E-mail: yuhy33@mail.sysu.edu.cn

**Abstract:** In moiré-patterned van der Waals structures of transition metal dichalcogenides, correlated insulators can form under integer and fractional fillings, whose transport properties are governed by various quasiparticle excitations including holons, doublons and interlayer exciton insulators. Here we theoretically investigate the nearest-neighbor inter-site hoppings of holons and interlayer exciton insulators. Our analysis indicates that these hopping strengths are significantly enhanced compared to that of a single carrier. The underlying mechanism can be attributed to the strong Coulomb interaction between carriers at different sites. For the interlayer exciton insulator consisting of a holon and a carrier in different layers, we have also obtained its effective Bohr radius and energy splitting between the ground and first-excited states.

In recent years, the formation of long-wavelength moiré superlattice patterns in van der Waals stacking of two-dimensional (2D) semiconducting transition metal dichalcogenides (TMDs) has emerged as a novel platform for studying exotic quantum states with vast tunability [1,2]. The moiré pattern offers an intrinsic superlattice system with a wavelength tunable through varying the interlayer twist angle or applying a heterostrain. Electrons and holes in the moiré pattern experience periodic modulating moiré potentials, which can give rise to superlattice mini bands for the carrier as well as moiré excitons with multiple optical resonances [3-8]. Flat mini bands have been realized in moiré systems with deep moiré potentials [9-12], where the carriers' kinetic energy is strongly suppressed. Combined with the enhanced Coulomb interaction from the layered geometry and weak dielectric screening of the surrounding environment, a series of correlated electronic states have been observed under specific doping densities, including the Mott insulators under integer fillings ($v = 1, 2, \ldots$) and generalized Wigner crystals under fractional fillings ($v = 1/3, 2/3, 1/2, \ldots$) [13-24]. These findings suggest that the moiré pattern can be an ideal system for studying novel correlated phenomena.

In the Mott insulator under $v = 1$, a pair of a doubly-occupied site in the upper Hubbard band and an empty site in the lower Hubbard band can be optically excited, which behaves similarly to the electron-hole pair in conventional semiconductors [25].

The empty (doubly-occupied) site corresponds to a novel quasiparticle termed as the holon (doublon), whose charge is opposite (the same) to that of the carrier. In generalized Wigner crystals under $v$ = 1/3, 2/3, 1/2, etc., the doublon cannot be directly obtained by adding a carrier to the system as not all sites are occupied, but removing a carrier from a Mott insulator or generalized Wigner crystal can always give rise to a holon. Similar to the carrier in the undoped limit ($v \to 0$), the holon can move in the system through the inter-site hopping thus introduces a finite conductance. In van der Waals multilayer systems, the carrier distribution between layers can be continuously tuned by an applied interlayer bias. When a carrier is moved from the Mott insulator in the moiré-patterned layer to the neighboring layer, it will bound with the resultant holon by the interlayer Coulomb potential to form an interlayer exciton insulator (IXI). A series of experiments have reported the observation of IXIs in Coulomb-coupled moiré-monolayer and double-moiré heterostructures [26-30]. An IXI can also move in the system through inter-site hopping, forming excitonic fluids and charge density waves. However, despite these interesting experimental progresses, fundamental properties of these IXIs are still not fully understood yet.

In this work, we focus on the Mott insulator under $v$ = 1 and generalized Wigner crystals under $v$ = 2/3 and 1/3, and analyze the inter-site hopping dynamics for the holon and IXI. Using a mean-field treatment on the interaction between carriers, we obtained hopping strengths $t_h^{(v=1)}$ and $t_h^{(v=2/3)}$ for holons under $v$ = 1 and $v$ = 2/3, respectively, as well as fundamental properties of the IXI under $v$ = 1 (including the effective Bohr radius, energy splitting between the ground and first-excited states and inter-site hopping strength $t_{IX}$). We find that, the hopping strengths $t_h^{(v=1)}$ and $t_h^{(v=2/3)}$ of holons and $t_{IX}$ of IXIs in TMDs moiré patterns are significantly larger than the single-particle hopping strength $t_s$ in the $v$ = 0 limit. This is found to be induced by the strong inter-site Coulomb repulsion. In contrast, in conventional Mott insulators with on-site repulsion only, the inter-site hopping strengths of holons and doublons are the same as the single-particle hopping [25].

We consider a van der Waals moiré-monolayer system where a moiré-patterned TMDs bilayer is separated from a TMDs monolayer by a sheet of few-layer hexagonal boron nitride (hBN). The moiré pattern has a wavelength $\lambda$, where carriers experience a moiré potential $V_{\text{moiré}}$. We assume $V_{\text{moiré}}$ to be in the following periodic form:

$$V_{\text{moiré}}(\boldsymbol{r}) = -\frac{2\Delta}{9} \sum_{j=1}^{3} \cos(\boldsymbol{b}_j \cdot \boldsymbol{r}). \qquad (1)$$

Here $\boldsymbol{b}_{1,2}$ are the primitive reciprocal lattice vectors of the moiré pattern and $\boldsymbol{b}_3 =$

$-\mathbf{b}_1 - \mathbf{b}_2$, see Fig. 1(a). $\Delta$ represents the modulation range of the moiré potential, which also characterizes the confinement strength of the moiré potential. Generally $\Delta$ varies with $\lambda$, but here we treat them as independent parameters. The potential landscape of $V_{\text{moiré}}$ is shown in Fig. 1(a), which shows ordered arrays of minima located at superlattice vectors denoted as $\mathbf{R}_n$. $V_{\text{moiré}}$ introduces a series of mini bands to the carrier, which can be obtained by diagonalizing the following Hamiltonian

$$\widehat{H}_q = \sum_G \frac{\hbar^2(\mathbf{q} + \mathbf{G})^2}{2m} |\mathbf{q} + \mathbf{G}\rangle\langle \mathbf{q} + \mathbf{G}| + \sum_{G,G'} V(\mathbf{G} - \mathbf{G}')|\mathbf{q} + \mathbf{G}\rangle\langle \mathbf{q} + \mathbf{G}'|. \quad (2)$$

Here $|\mathbf{q} + \mathbf{G}\rangle$ represents the monolayer Bloch state with a wave vector $\mathbf{q} + \mathbf{G}$, with $\mathbf{q}$ restricted in the mini Brillouin zone and $\mathbf{G}$ the reciprocal lattice vector of the moiré superlattice pattern. $m \approx 0.5 m_0$ is the effective mass of the monolayer Bloch state with $m_0$ the free electron mass, $V(\mathbf{G})$ is the Fourier coefficient of $V_{\text{moiré}}$.

Under a strong modulation of the moiré potential, the lowest mini band can be well described by a single-orbital tight-binding model, with the orbital basis being gaussian wavepackets localized at potential minima (see the illustration in Fig. 1(a)). The single-particle inter-site hopping $t_s$ between nearest-neighbor sites, which determines the band width, can be obtained by fitting the tight-binding band to that from diagonalizing Eq. (2). The obtained $t_s$ values as functions of the moiré wavelength $\lambda$ under several different values of $\Delta$ are shown in Fig. 1(b). Generally, $t_s$ falls in the order of $O(0.1)$ to $O(1)$ meV and decays exponentially with $\lambda$, which is due to the fact that the ratio $\sigma_0/\lambda = \sqrt{\frac{3\hbar}{4\pi\lambda}}(m\Delta)^{-\frac{1}{4}}$ between the wavepacket width $\sigma_0$ and nearest-neighbor distance $\lambda$ decreases with $\lambda$. Meanwhile, a strong moiré confinement $\Delta$ narrows the wavepacket width $\sigma_0$ thus also decreases $t_s$.

The many-body interaction effect becomes more and more important with the increase of the carrier density. Under a density of one carrier per moiré supercell (i.e., a filling factor of $v = 1$), the system becomes a Mott insulator when $t_s$ is much weaker than the on-site Coulomb repulsion between two wavepackets located at the same moiré potential minimum [13-17,22]. Furthermore, when the inter-site Coulomb repulsion is much larger than $t_s$, the system can become a generalized Wigner crystal under certain fractional filing factors ($v = $ 1/3, 1/2, 2/3, etc.) which breaks the discrete translational symmetry of the moiré superlattice [14,16,18-20,24]. The experimentally measured on-site and nearest-neighbor inter-site repulsions are found to be ~ 100 meV and several tens meV, respectively [22,31,32], both are one to two orders of magnitude larger than our calculated $t_s$.

Fig. 1(c) illustrates a Mott insulator under a filling factor $v = 1$. Here we model the correlated insulator system by carriers in wavepacket forms localized at $\boldsymbol{R}_n$, with a wavepacket width $2\sigma$ determined by both the moiré confinement strength $\Delta$ and the inter-site Coulomb interaction [33]. To evaluate the value of $\sigma$, we consider the carrier localized at $\boldsymbol{R}_0 \equiv 0$, and write the total potential it experiences as

$$V_{\text{Mott}}(\boldsymbol{r}) = \sum_{n \neq 0} \int d\boldsymbol{r}_n \rho(\boldsymbol{R}_n + \boldsymbol{r}_n) V_{\text{intra}}(\boldsymbol{R}_n + \boldsymbol{r}_n - \boldsymbol{r}) + V_{\text{moiré}}(\boldsymbol{r}). \tag{3}$$

Here $\rho(\boldsymbol{R}_n + \boldsymbol{r}_n) \equiv \frac{1}{\pi \sigma^2} e^{-r_n^2/\sigma^2}$ is the density distribution of the carrier localized at the $n$-th superlattice vector $\boldsymbol{R}_n$. $V_{\text{intra}}(\boldsymbol{R}_n + \boldsymbol{r}_n - \boldsymbol{r})$ corresponds to the intralayer Coulomb interaction between two carriers located at $\boldsymbol{R}_n + \boldsymbol{r}_n$ and $\boldsymbol{r}$, respectively. In layered semiconductor systems, the Coulomb interaction is affected by the 2D screening of the layered materials as well as the dielectric constant $\epsilon$ of the surrounding environment. We use the following form of $V_{\text{intra}}(\boldsymbol{r})$ from Ref. [34]:

$$V_{\text{intra}}(\boldsymbol{r}) = \frac{1}{\epsilon} \int_0^\infty dk \frac{[1 + kr_0(1 - e^{-2kD})]J_0(kr)}{[1 + kr_0(1 - e^{-kD})][1 + kr_0(1 + e^{-kD})]}. \tag{4}$$

Here $J_0$ is the Bessel functions of the first kind, and $r_0 \approx 4.5/\epsilon$ nm corresponds to the screening length of monolayer TMDs [35]. Note that in the $D \to \infty$ limit, $V_{\text{intra}}(\boldsymbol{r}) = \frac{1}{\epsilon} \int_0^\infty dk \frac{J_0(kr)}{1+kr_0}$ becomes the widely adopted Keldysh form of the Coulomb potential [36-38]. Here we set $D = 1$ nm as the value in experiments, and $\epsilon = 4.5$ as the average dielectric constant of thick hBN encapsulation layers. The resultant $V_{\text{Mott}}(\boldsymbol{r})$ shows a deep single-well form, see Fig. 1(d). Near the potential minimum, $V_{\text{Mott}}(\boldsymbol{r}) \approx V_{\text{Mott}}(0) + \frac{r^2}{2} \left( \frac{16\pi^2 \Delta}{9\lambda^2} + \sum_{n \neq 0} \int d\boldsymbol{r}_n \rho(\boldsymbol{R}_n + \boldsymbol{r}_n) \frac{\partial^2 V(\boldsymbol{R}_n + \boldsymbol{r}_n)}{\partial r_n^2} \right)$ can be well approximated by a harmonic potential, and the width $2\sigma$ of the ground state wavepacket can be calculated self-consistently. The obtained values of $\sigma$ as functions of $\lambda$ are indicated in Fig. 1(e). For the considered range of $\lambda$, there is $\sigma \ll \lambda$ when $\Delta$ is large. $\sigma$ is also smaller than the single-particle wavepacket width $\sigma_0$, showing the significance of inter-site Coulomb repulsions. Both $\sigma$ and $\sigma_0$ fall in the order of nm. Such small carrier spatial distributions lead to the strong on-site Coulomb repulsion.

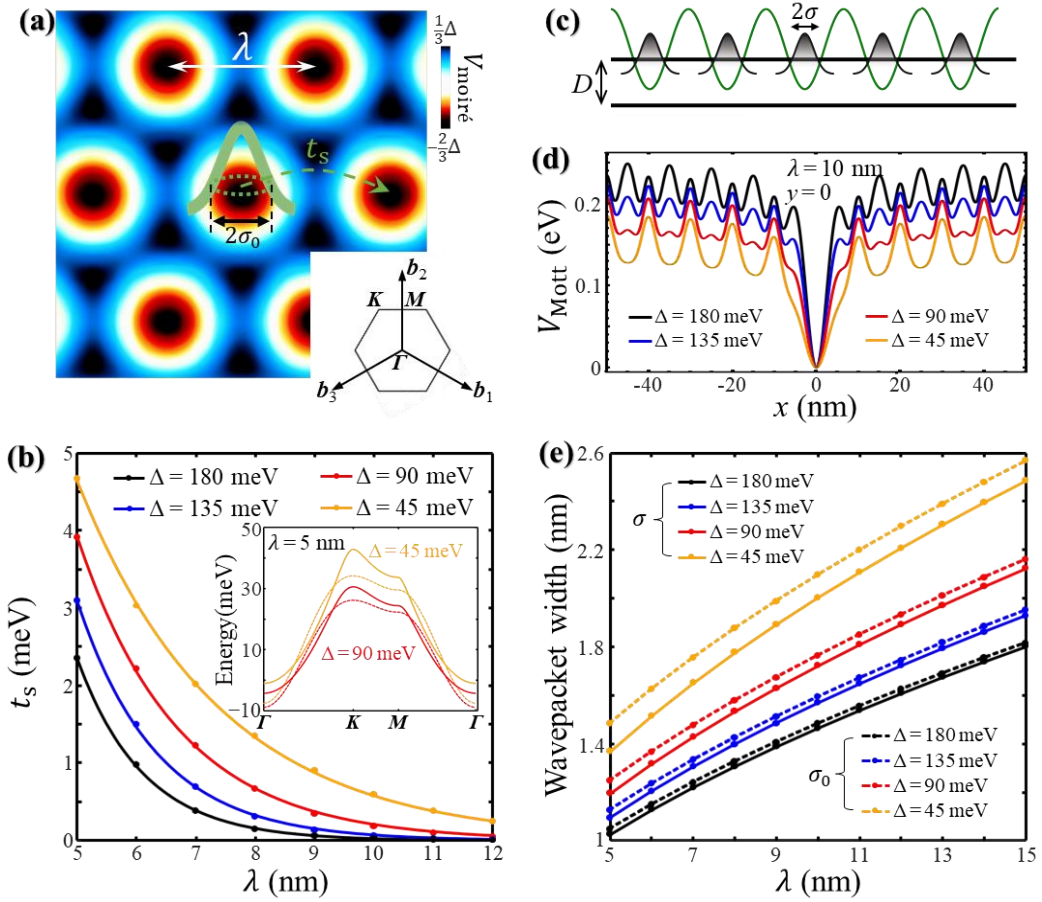

**Fig. 1** (a) The landscape of the moiré potential given by Eq. (1), with $\lambda$ the moiré wavelength and $\Delta$ the potential modulation range. A carrier wavepacket localized at a potential minimum can hop to its nearest-neighbor site with a hopping strength $t_s$. $2\sigma_0$ is the wavepacket width. (b) The obtained single-particle hopping $t_s$ (dots) as a function of $\lambda$, and the fittings with exponential decay functions (lines). The inset indicates the numerically solved lowest bands (solid lines) and their fittings using a tight-binding model (dashed lines) under $\lambda = 5$ nm, $\Delta = 45$ and 90 meV. (c) A schematic illustration of the Mott insulator under $v = 1$ in the upper-layer. Each carrier is modeled by a gaussian wavepacket with width $2\sigma$. There is no carrier in the lower-layer separated by a vertical distance $D$ from the upper-layer. (d) The total potential $V_{\text{Mott}}$ experienced by a carrier in the upper-layer Mott insulator, given by the sum of the moiré potential and the inter-site Coulomb interactions. The curves are plotted along $y = 0$. (e) The calculated wavepacket widths $\sigma$ (solid lines) and $\sigma_0$ (dashed lines) as a function of $\lambda$. The lines are guides to the eye. Under a given $\Delta$, $\sigma$ in the $v = 1$ Mott insulator is smaller than the single-particle width $\sigma_0$.

After removing a carrier from the fully filled ($v = 1$) Mott insulator, the resultant empty site corresponds to a holon which can freely move in the 2D plane. The holon inter-site hopping is equivalent to the hopping of a nearest-neighbor carrier to the empty site, see Fig. 2(a). To evaluate the holon hopping strength $t_h^{(v=1)}$, we focus on the carrier in the dashed box of Fig. 2(a). It experiences a mean-field confinement potential

$$V_{\text{upper}}^{(\nu=1)}(\mathbf{r}) = \sum_{n \neq 0,1} \int d\mathbf{r}_n \rho(\mathbf{R}_n + \mathbf{r}_n) V_{\text{intra}}(\mathbf{R}_n + \mathbf{r}_n - \mathbf{r}) + V_{\text{moiré}}(\mathbf{r}). \tag{5}$$

Here the sum $\sum_{n \neq 0,1}$ is over all carriers outside the dashed box of Fig. 2(a), which are approximated by wavepackets with width $2\sigma$ given in Fig. 1(e). The obtained potential $V_{\text{upper}}^{(\nu=1)}(\mathbf{r})$ shows a double-well shape with the two minima located close to the moiré potential minima, see Fig. 2(b). The two lowest-energy states of a carrier in such a double-well potential can be viewed as the symmetric and antisymmetric superpositions of the two states localized at the two potential minima, whose energy splitting corresponds to $2t_h^{(\nu=1)}$. By numerically solving the two lowest-energy states under the given $V_{\text{upper}}^{(\nu=1)}(\mathbf{r})$, we obtain the value of $t_h^{(\nu=1)}$ which decays exponentially with $\lambda$, see Fig. 2(c). We note that $t_h^{(\nu=1)}$ is significantly enhanced compared to $t_s$ (see Fig. 2(c) inset for the ratio $t_h^{(\nu=1)}/t_s$), which can be understood from the following consideration. When an empty site is introduced to the system, its nearby carriers experience non-zero total Coulomb forces which then displace their center positions away from the moiré potential minima. As a result, the in-plane distance between the two minima of $V_{\text{upper}}^{(\nu=1)}(\mathbf{r})$ becomes smaller than $\lambda$, which thus enhance the nearest-neighbor hopping strength. We emphasize that such an enhancement is due to the strong inter-site repulsion which comes from the long-range nature of the Coulomb interaction. In a standard Hubbard model with on-site interactions only, $t_h^{(\nu=1)}$ is expected to be the same as $t_s$.

The same analysis can be applied to the generalized Wigner crystal formed under $\nu = 2/3$, where a holon is also obtained after removing a carrier, see Fig. 2(d). For the carrier in the dashed box of Fig. 2(d), the potential $V_{\text{upper}}^{(\nu=2/3)}(\mathbf{r})$ it experienced is shown in Fig. 2(e). Compared to $V_{\text{upper}}^{(\nu=1)}(\mathbf{r})$ in Fig. 2(b), a series of local potential minima emerge in $V_{\text{upper}}^{(\nu=2/3)}(\mathbf{r})$ which correspond to unoccupied moiré potential minima in the original $\nu = 2/3$ generalized Wigner crystal. Again, we numerically solve the two lowest-energy states under the effect of $V_{\text{upper}}^{(\nu=2/3)}(\mathbf{r})$ to get the inter-site hopping $t_h^{(\nu=2/3)}$, see Fig. 2(f). We find $t_h^{(\nu=2/3)} > t_h^{(\nu=1)}$. Note that in the $\nu = 2/3$ generalized

Wigner crystal, a carrier near the holon also experiences a non-zero total Coulomb force whose magnitude is almost the same as that of the $v = 1$ case [39]. Compared to $t_h^{(v=1)}$, the larger magnitude of $t_h^{(v=2/3)}$ is thus induced by the weaker confinement strengths, which leads to a larger wavepacket extension $\sigma$ and larger displacements away from the moiré potential minima.

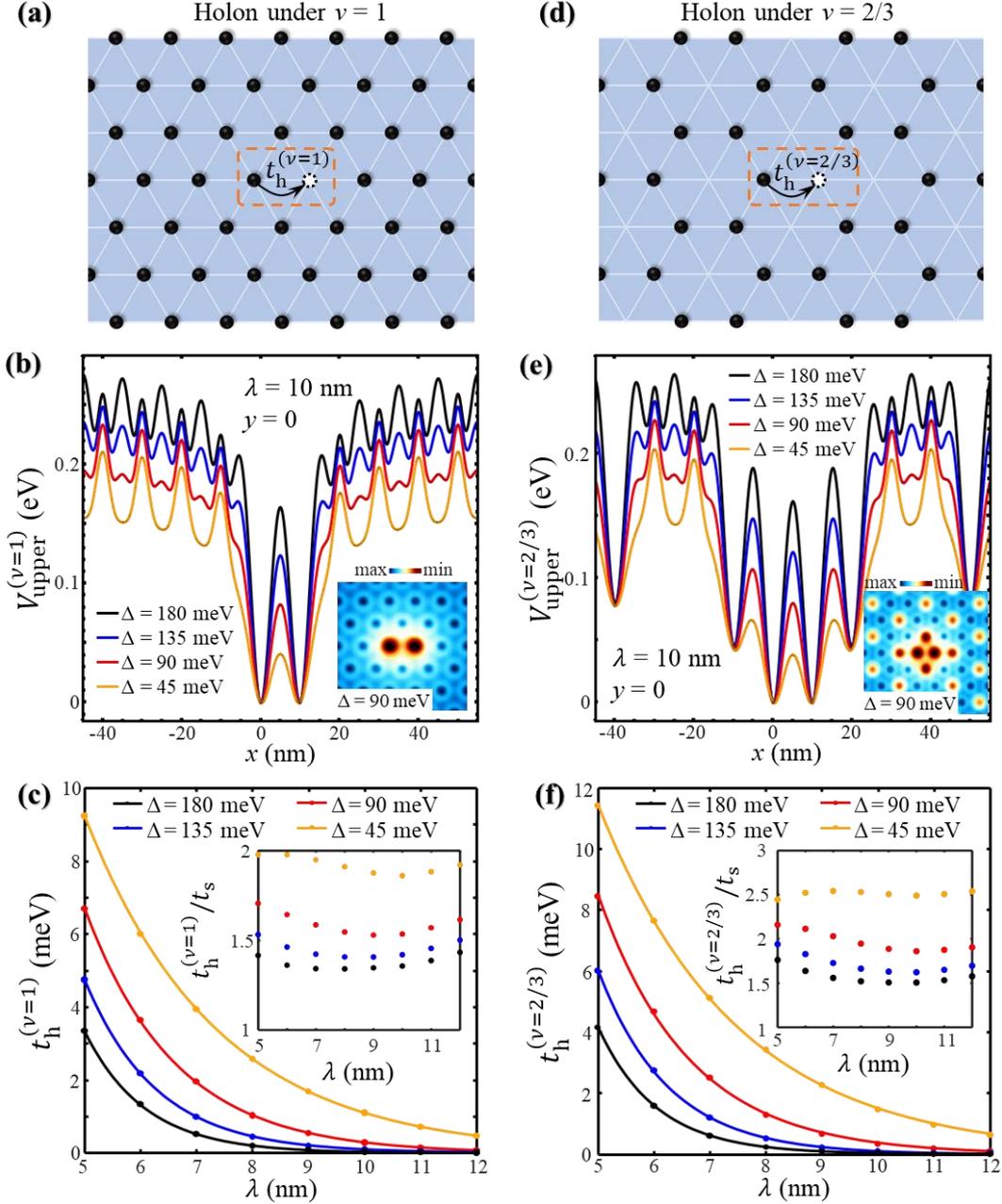

**Fig. 2** (a) The holon (the dashed empty circle) obtained by removing a carrier in the Mott insulator under $v = 1$. The nearest-neighbor inter-site hopping $t_h^{(v=1)}$ of the holon is equivalent to the hopping of a carrier to the empty site. (b) The total potential $V_{\text{upper}}^{(v=1)}$ experienced by the carrier in the dashed box of (a), where the minimum potential value is set as 0. The lines are obtained along $y = 0$, and the inset shows the 2D landscape of $V_{\text{upper}}^{(v=1)}$ under $\Delta = 90$ meV. (c) The numerically solved $t_h^{(v=1)}$ (dots) as a function of $\lambda$, and the fittings with exponential decay functions (lines). The inset shows the ratio $t_h^{(\nu=1)}/t_s$. (d) A holon obtained after removing a carrier in the honeycomb-type generalized Wigner crystal under $\nu = 2/3$, and its inter-site hopping $t_h^{(\nu=2/3)}$. (e) The potential $V_{\text{upper}}^{(\nu=2/3)}$ experienced by the carrier in the dashed box of (d). (f) The numerically solved $t_h^{(\nu=2/3)}$. The inset shows the ratio $t_h^{(\nu=2/3)}/t_s$.

Applying an interlayer bias can distribute carriers between the two TMDs layers. When a carrier in the upper-layer Mott insulator ($\nu = 1$) is moved to the lower-layer, it will be bound to the empty site in the upper-layer due to the interlayer Coulomb interaction, see Fig. 3(a). The resultant bound state consisting of a carrier and a holon in different layers corresponds to an IXI, which can also move in the bilayer plane through the inter-site hopping. The total interlayer Coulomb potential experienced by the carrier in the lower-layer has the following form:

$$V_{\text{IX}}(\boldsymbol{r}) = \sum_{n \neq 0} \int d\boldsymbol{r}_n \rho(\boldsymbol{R}_n + \boldsymbol{r}_n) V_{\text{inter}}(\boldsymbol{R}_n + \boldsymbol{r}_n - \boldsymbol{r}). \tag{6}$$

Here $V_{\text{inter}}(\boldsymbol{R}_n + \boldsymbol{r}_n - \boldsymbol{r})$ corresponds to the interlayer Coulomb interaction between a carrier in the upper-layer with an in-plane position $\boldsymbol{R}_n + \boldsymbol{r}_n$ and a carrier in the lower-layer with an in-plane position $\boldsymbol{r}$. We use the following form of $V_{\text{inter}}(\boldsymbol{r})$ from Ref. [34]:

$$V_{\text{inter}}(\boldsymbol{r}) = \frac{1}{\epsilon} \int_0^\infty dk \frac{e^{-kD} J_0(kr)}{[1 + kr_0(1 - e^{-kD})][1 + kr_0(1 + e^{-kD})]}. \tag{7}$$

Fig. 3(b) shows the landscape of $V_{\text{IX}}(\boldsymbol{r})$ which is also in a single-well form. Due to the absence of the moiré potential in the lower-layer and the weaker strength of $V_{\text{inter}}(\boldsymbol{r})$ than $V_{\text{intra}}(\boldsymbol{r})$, $V_{\text{IX}}(\boldsymbol{r})$ has a shallower depth and slower rise speed compared to $V_{\text{Mott}}(\boldsymbol{r})$ when moving away from $\boldsymbol{r} = 0$. Meanwhile, under a fixed $\lambda$ value, the moiré potential modulation range $\Delta$ in the upper-layer, which changes the upper-layer carrier wavepacket width $2\sigma$, barely affects the shape of $V_{\text{IX}}(\boldsymbol{r})$ near $\boldsymbol{r} = 0$. Since the minimum of $V_{\text{IX}}(\boldsymbol{r})$ has an in-plane position located at the empty site, the carrier in the lower-layer will be trapped at the holon site in the upper-layer, giving rise to an IXI. By approximating $V_{\text{IX}}(\boldsymbol{r})$ near $\boldsymbol{r} = 0$ as a harmonic potential, the low-energy states of IXI can be described by a 2D harmonic oscillator model. Fig. 3(c) shows the numerically solved ground state energy $E_0$ and first excited state energy $E_1$. Generally, $E_0$ and $E_1$ satisfy $E_1 \approx 2E_0$ and are on the order of $O(10)$ meV, which rise with the reduction of $\lambda$. The IXI ground state corresponds to a lower-layer carrier in the gaussian wavepacket form, whose width $\sigma'$ as a function of $\lambda$ is shown in Fig. 3(d). $\sigma'$ is found to be significantly larger than the width $\sigma$ of upper-layer carriers given in Fig. 1(e), but still much smaller than $\lambda$. We note that $\sigma'$ corresponds to the root-mean-square distance between the holon and carrier in the IXI, which can be viewed as an effective Bohr radius. Compared to the Bohr radius $a_B$ of the optically excited interlayer exciton in bilayer TMDs which is ~ 2 nm measured from experiments [40,41], the value of $\sigma'$ is comparable to $a_B$. However, $\sigma'$ increases linearly with $\lambda$ whereas $a_B$ is insensitive to $\lambda$.

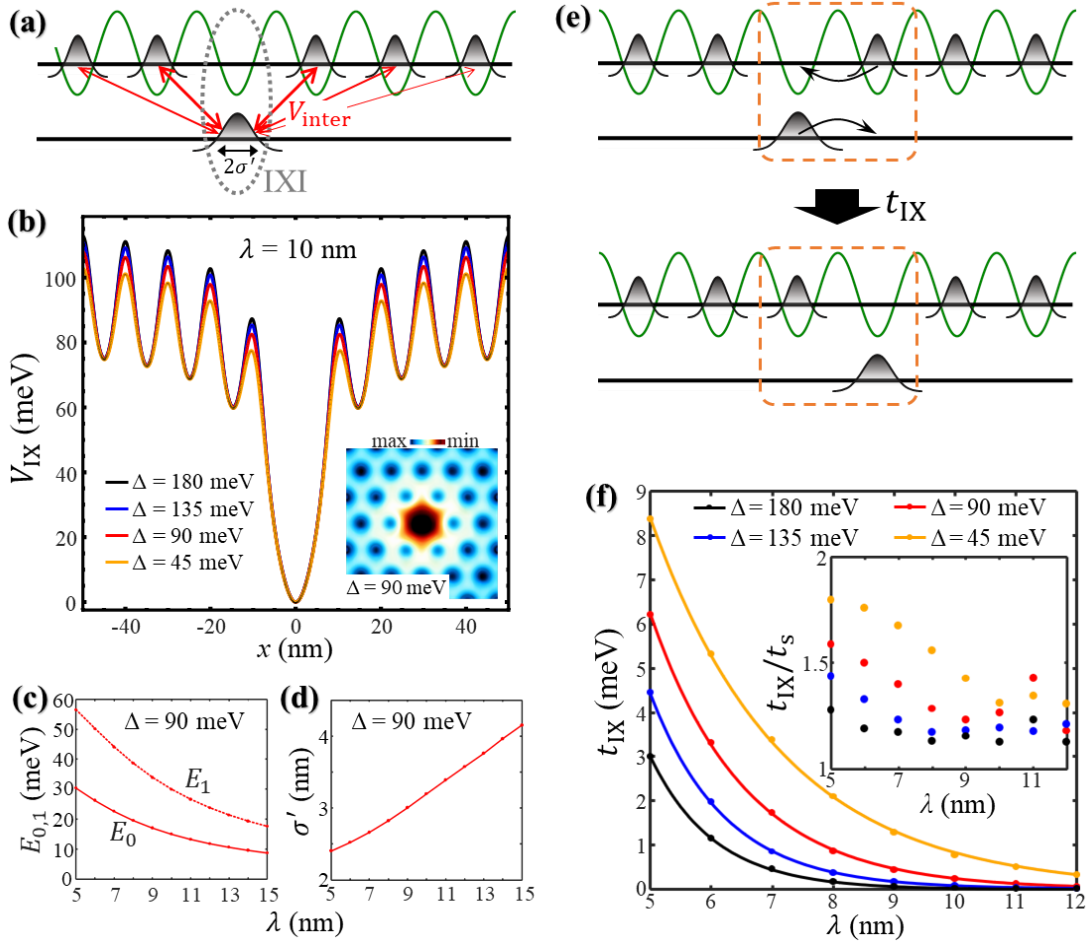

**Figure 3** (a) The interlayer exciton insulator (IXI) formed by a holon in the upper-layer Mott insulator and a carrier in the lower-layer. Due to the interlayer Coulomb repulsion $V_{\text{inter}}$, the lower-layer carrier is in a wavepacket form with a width $2\sigma'$. (b) The potential $V_{\text{IX}}$ experienced by the lower-layer carrier, given by the total interlayer Coulomb potential from upper-layer carriers. (c) The IXI ground state energy $E_0$ and first excited state energy $E_1$ as functions of $\lambda$ under $\Delta = 90$ meV. (d) The wavepacket width $\sigma'$ of the lower-layer carrier under $\Delta = 90$ meV. (e) A schematic illustration for the inter-site hopping process of the IXI. (f) The numerically solved inter-site hopping strength $t_{\text{IX}}$ (dots) as a function of $\lambda$, and the fitting with exponential decay functions (lines). The inset shows the ratio $t_{\text{IX}}^{(\text{fit})}/t_s$.

To analyze the inter-site hopping $t_{\text{IX}}$ of the IXI, we consider the two carriers in the dashed box of Fig. 3(e) and treat the other carriers as the background charge density. The corresponding two-particle Hamiltonian is

$$\hat{H} = \hat{H}_{\text{upper}} + \hat{H}_{\text{lower}} + V_{\text{inter}}(\boldsymbol{r} - \boldsymbol{r}'),$$

$$\hat{H}_{\text{upper}} = -\frac{\hbar^2}{2m}\frac{\partial^2}{\partial r^2} + V_{\text{upper}}^{(\nu=1)}(\boldsymbol{r}), \quad (8)$$

$$\hat{H}_{\text{lower}} = -\frac{\hbar^2}{2m}\frac{\partial^2}{\partial r'^2} + V_{\text{lower}}^{(\nu=1)}(\boldsymbol{r}').$$

Here $\hat{H}_{\text{upper}}$ is the Hamiltonian of the upper-layer carrier, with the double-well potential $V_{\text{upper}}^{(\nu=1)}(\boldsymbol{r})$ given by Eq. (5) and Fig. 2(b). $\hat{H}_{\text{lower}}$ is the Hamiltonian of the lower-layer carrier, with $V_{\text{lower}}^{(\nu=1)}(\boldsymbol{r}') = \sum_{n\neq 0,1}\int d\boldsymbol{r}_n \rho(\boldsymbol{R}_n + \boldsymbol{r}_n)V_{\text{inter}}(\boldsymbol{R}_n + \boldsymbol{r}_n - \boldsymbol{r}')$ induced by the interlayer Coulomb repulsion from background carriers outside the dashed box of Fig. 3(e). A set of tensor-product basis $\psi_{\text{upper}}(\boldsymbol{r}) \otimes \psi_{\text{lower}}(\boldsymbol{r}')$ can be constructed from the eigenfunctions $\psi_{\text{upper}}(\boldsymbol{r})$ of $\hat{H}_{\text{upper}}$ and $\psi_{\text{lower}}(\boldsymbol{r}')$ of $\hat{H}_{\text{lower}}$, which is used to diagonalize the two-particle Hamiltonian $\hat{H}$. The energy splitting between the lowest-energy and second-lowest-energy states then gives $2t_{\text{IX}}$, with the obtained values indicated in Fig. 3(f). Being a two-particle process, the strength of $t_{\text{IX}}$ is weaker than $t_h^{(\nu=1)}$ and $t_h^{(\nu=2/3)}$ of holons, but still larger than the single-particle hopping $t_s$ (see Fig. 3(f) inset for $t_{\text{IX}}/t_s$). For a better comparison between the above discussed various hopping strengths, we plot $t_s$, $t_{\text{IX}}$, $t_h^{(\nu=1)}$, and $t_h^{(\nu=2/3)}$ in the log scale under different $\Delta$ values in Fig. A1 of the Appendix. For all the considered $\Delta$ values, there is $t_s < t_{\text{IX}} < t_h^{(\nu=1)} < t_h^{(\nu=2/3)}$. All four hopping strengths decay exponentially with $\lambda$, with nearly the same decay lengths under a given $\Delta$. The same analysis can be applied to IXIs formed in the $\nu = 2/3$ generalized Wigner crystal, where the interlayer Coulomb potential $V_{\text{IX}}^{(\nu=2/3)}(\boldsymbol{r})$ is shallower than that of the $\nu = 1$ case in Fig. 3(b). This then leads to a smaller splitting $E_1 - E_0$ between the first excited and ground states and larger wavepacket extension $\sigma'$ of the lower-layer carrier. As a result, the inter-site hopping of IXIs under $\nu = 2/3$ case should have a larger strength than $t_{\text{IX}}$ under $\nu = 1$. However, the IXI under $\nu = 2/3$ is more difficult to observe in experiments due to its weaker binding energy.

Finally, we investigate the inter-site hopping $t_h^{(\nu=1/3)}$ of holons in the triangular-type generalized Wigner crystal under $\nu = 1/3$ (see Fig. 4(a)). Following the same procedure for the generalized Wigner crystal under $\nu = 2/3$, we consider the carrier in the dashed box of Fig. 4(a) and plot the potential it experienced in Fig. 4(b). Fig. 4(b) is obtained under $\Delta = 90$ meV and $\lambda = 5$ nm, whereas using other parameter values only quantitative changes the potential landscape. Besides the global minima located at the carrier site and the holon site, a series of local minima with potentials rather close to the global minima emerge in the dashed box, which correspond to unoccupied sites of the $\nu = 1/3$ generalized Wigner crystal. From the potential in Fig. 4(b), the eigenstates and the corresponding energies of the carrier in the dashed box can be numerically solved. The energy splitting between the two lowest-energy states gives $2t_h^{(\nu=1/3)}$, whose values as functions of $\lambda$ are given in Fig. 4(c). We note that $t_h^{(\nu=1/3)}$ corresponds to a hopping process between two next-nearest-neighbor sites (see Fig. 4(a)), thus is expected to be exponentially weak compared to the previously discussed nearest-neighbor hopping processes. However, the obtained ratio $t_h^{(\nu=1/3)}/t_s$ is smaller than 1 and decays exponentially with $\lambda$ only when $\lambda$ is above a certain threshold (see Fig. 4(c) inset). This threshold value decreases with the increase of $\Delta$. For $\lambda$ smaller than this threshold, $t_h^{(\nu=1/3)}$ can be comparable to $t_s$. To investigate the underlying mechanism, we consider the wave functions $\psi_0(\mathbf{r})$ and $\psi_1(\mathbf{r})$ of the lowest-energy and second-lowest-energy states, respectively, which are symmetric and antisymmetric after exchanging the two global minimum sites. Fig. 4(d) shows the wave function intensities $|\psi_0(\mathbf{r})|^2$ and $|\psi_1(\mathbf{r})|^2$ under the potential in Fig. 4(b) where $\Delta = 90$ meV and $\lambda = 5$ nm is below the corresponding threshold. $|\psi_0(\mathbf{r})|^2$ has significant distributions at local minima between the two global minima, implying that the inter-site Coulomb interaction can push the carrier into originally unoccupied sites with higher energies and $t_h^{(\nu=1/3)}$ can be largely mediated by the nearest-neighbor hopping. However, the distribution at local minima vanishes when increasing $\Delta$ or $\lambda$. This then explains why the ratio $t_h^{(\nu=1/3)}/t_s$ is close to 1 for $\lambda$ below the threshold, but becomes exponentially small when $\lambda$ is above the threshold.

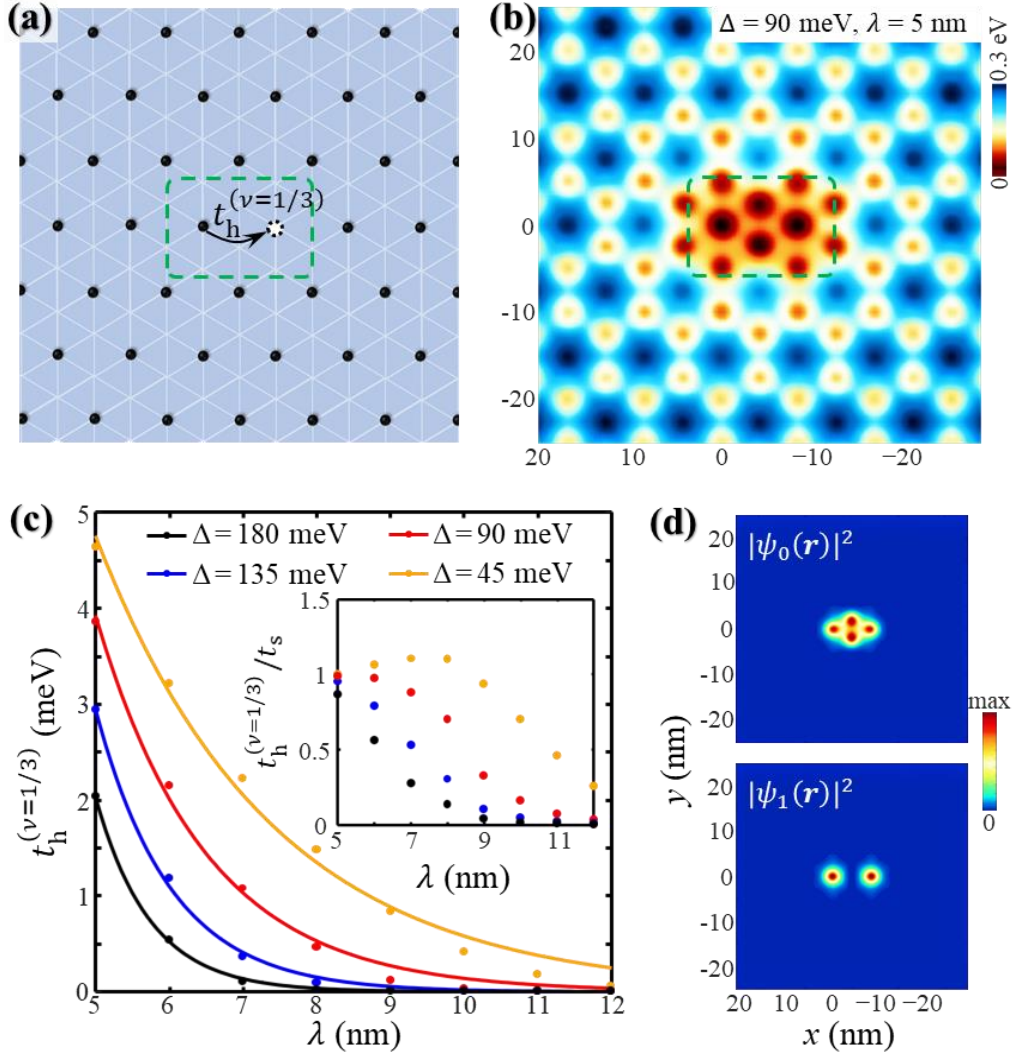

**Figure 4**. (a) A holon (the dashed empty circle) in the generalized Wigner crystal under $v = 1/3$, and its inter-site hopping $t_h^{(v=1/3)}$. (b) The potential experienced by the carrier in the dashed box in (a) under $\Delta = 90$ meV and $\lambda = 5$ nm, which is induced by the sum of Coulomb interactions from all other carriers outside the dashed box. (c) The numerically solved $t_h^{(v=1/3)}$ (dots) as a function of $\lambda$, and the fitting with exponential decay functions (lines). The inset shows the ratio $t_h^{(v=1/3)}/t_s$. (d) The numerically obtained wave function intensities $|\psi_0(r)|^2$ and $|\psi_1(r)|^2$ of the lowest-energy and second-lowest-energy states, respectively, for the carrier in the dashed box under the potential in (b).

In summary, we have theoretically investigated the inter-site hopping strengths of holons and IXIs in correlated insulators of moiré patterned TMDs van der Waals structures. The calculated hopping strengths of holons are significantly larger than the single-particle hopping of a carrier in the moiré superlattice pattern. Such a behavior is found to originate from the strong inter-site Coulomb repulsion between carriers. Although the inter-site hopping of the IXI corresponds to a two-particle process, its strength is also found to be larger than the single-particle one, again showing the enhancement by the interaction effect. We have also obtained several fundamental properties of the IXI, including its effective Bohr radius and energy splitting between the ground and first-excited states which haven't been investigated in earlier literatures. Our studies can provide insights for understanding transport properties of various quasiparticles emerged in correlated insulators of TMDs moiré patterns.

**Acknowledgement.** H.Y. acknowledges support by NSFC under grant No. 12274477, and the Department of Science and Technology of Guangdong Province in China (2019QN01X061).

## Appendix: comparisons between various inter-site hopping strengths

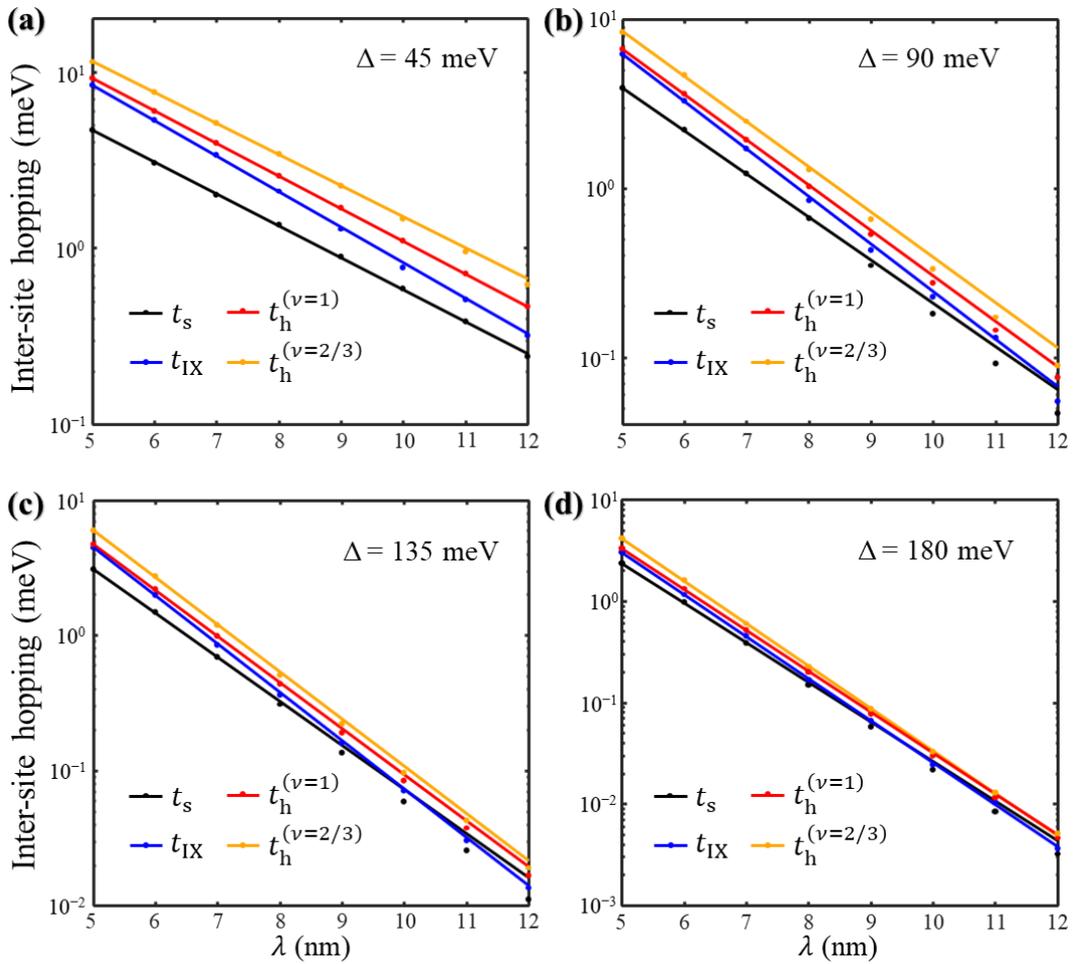

**Figure A1**. A comparison between inter-site hopping strengths of the single-particle ($t_s$), IXI ($t_{IX}$), holon in the Mott insulator under $v = 1$ ($t_h^{(v=1)}$) and holon in the generalized Wigner crystal under $v = 2/3$ ($t_h^{(v=2/3)}$). The dots are numerical results, and lines are fittings using exponential decay functions.